%% file: main.tex
\documentclass[a4paper,11pt]{article}
\usepackage{aaskaiid}
\usepackage{graphicx} 
\usepackage{orcidlink}
\title{The science of the cycle of matter in our Galaxy with the SKA}
\author[1]{Marc Audard\orcidlink{0000-0003-4721-034X}}
\author[2]{Ke Wang\orcidlink{0000-0002-7237-3856}}

\affiliation[1]{Department of Astronomy, University of Geneva, Ch. Pegasi 51, 1290 Versoix, Switzerland}
\emailAdd{Marc.Audard@unige.ch}

\affiliation[2]{Kavli Institute for Astronomy and Astrophysics, Peking University, 5 Yiheyuan Road, Haidian District, Beijing 100871, People's Republic of China\\
}
\emailAdd{kwang.astro@pku.edu.cn}

\abstract{Exploring how matter cycles through the Galaxy, from the birth of stars in dense interstellar clouds to the ejection of matter and energy during a star’s final stages-requires a multi-faceted approach. Radio observations are essential to reveal the intricate interactions at play. By studying our Galaxy in detail, we can use it as a model to better understand these processes in galaxies as well. The Square Kilometre Array offers unique capabilities in wide-field, high-sensitivity, high-resolution spectroscopy and precise astrometry to revolutionise Galactic astrophysics. This chapter presents an overview of the science addressed in detail in chapters pertinent to the SKAO Science Working Group \textit{Our Galaxy}.
}

\date{July 2026}

\begin{document}

\maketitle

\section{Introduction}

The SKAO Science Working (SWG) \textit{Our Galaxy} (herafter OG) has the scientific aim to study the cycle of matter our Galaxy, the Milky Way. The SWG membership is, thus, very diverse, as the members study several aspects, from star formation in the interstellar medium (ISM) and molecular clouds to the late phase of a star losing matter and energy into the ISM, through magnetic activity, etc. The SKAO will help our community address many aspects of this ISM cycle in our Galaxy, thanks to its large field of view combined with high sensitivity and spectral resolution. Consequently, the various scientific topics pertaining to SKAO use both the SKA-Mid and SKA-Low arrays, albeit with different observational approaches.

Several chapters are presented here under the umbrella of the SWG OG. We provide below a short overview of the main scientific points addressed in each chapter. Those are addressed preferentially with SKA-Mid, since the scientific topics in OG are better resolved at higher frequencies, but SKA-Low offers additional opportunities as well. In addition, we should emphasise that some chapters overlap and offer synergies, and actual observing programs will take into account these aspects.

\section{Probing the different phases of the ISM}

\citet{Yamamoto01.2026.SKA} presented a discussion of the cold and neutral medium (CNM) in the ISM using the SKA1-Mid, with the goal to understand the physical processes taking place in the CNM, in particular in the early phase of molecular cloud formation. Their chapter shows that the combination of angular resolution (at arcsecond level) together with spectral resolution and high surface brightness sensitivty is a key feature of SKAO to study the sizes, morphology, and dynamics in the CNM, in particular at the scales of 1000 au, below the typical size of the field length associated with thermal instability (0.01 pc). They propose to take the advantage of the full polarisation capability of SKA1-Mid to measure the magnetic field via Zeeman measurements (see also Bourke et al.). Such measurements will determine the importance of magnetic fields in the star formation process. They emphasise that the SKA1-Mid capabilities will be decisive to test stability conditions and the two-phase turbulence model in the ISM, with the ultimate goal to better understand the transition from the multi-phase atomic ISM to molecular clouds.

\citet{Salas01.2026.SKA} focused their chapter on the diffuse (cold neutral and warm ionised) ISM by means of radiative recombination lines (RRL) in the radio. The RRL provide extinction free probes of the gas kinematics, density, temperature, and radiation fields in the warm ionised medium (WIM), a major component of the Milky Way, filling a large fraction of the Galaxy, and accounting for 90\% of the ionised hydrogen in the Galaxy, in addition to the CNM (see above). Different elements can be used as tracers, such as hydrogen or helium for ionised regions, or carbon RRLS to probe the cold gas in the layers where atomic hydrogen and molecular hydrogen transition. Again, the combination of the spatial, spectral, and wide-area capability of SKAO (both Low and Mid arrays) is a key ingredient to study this phase of the ISM with RRLs, and directly measure the thermal and turbulence balance, chemical enrichment, sizes, and ionisation rates at all scales down to arcseconds.

\citet{Traficante01.2026.SKA} have proposed a survey of the Galactic plane in the radio continuum band of SKA-Mid. The survey, of $630^2$ deg coverage ($-180\deg<l<30\deg$, $|b|\leq 1.5\deg$) would make use of simultaneous observations to cover 10-15 GHz nominally.
The survey would reach deep sensitivities with high angular resolution (smaller than 0.05 pc at distance up to 20 kpc), enabling the resolution of the earliest stages of ultracompact H~\textsc{ii} regions, or characterise jets up to 10 kpc away, probe planetary nebulae, evolved massive stars, and supernovae remnants, etc. Such a Galactic survey would provide a detailed view of star formation and evolution in our Galaxy, offering synergies with surveys at other wavelengths.

\citet{Karska01.2026.SKA} recognise the power of spectral surveys with SKAO and describe potential surveys, together with their astrophysical impact, in specific lines. For example, a Galactic plane survey in OH absorption, coupled with a similar survey in atomic hydrogen, would provide information on on matched pairs, and thus trace the molecular gas in the CNM and at different galactic radii. It would help us constrain the thermal and physical conditions in each component in both atomic and molecular gas, including the CO-dark diffuse molecular hydrogen clouds. The authors also suggest targeted observations toward H~\textsc{ii} regions using H$\alpha$ RRLs in the 1-15 GHz range, or observations of formaldehyde in absorption from Galactic clouds to trace the molecular gas masses and densities.

\citet{Schoedel01.2026.SKA} focus  specifically on the Galactic center, covering an area of $2.0\deg \times 0.4\deg$, i.e.,  $290 $ pc $ \times 60$ pc, centered on Sgr A$^*$ using SKA-Mid in Bands 2, 5a and 5b (1.36, 6.55, and 11.85 GHz). Compared to the much wider and high-frequency survey of Traficante et al., the Galactic center survey would go much deeper, albeit with a much smaller area. However, it will probe star formation and evolution in the extreme environments of the central molecular zone, with angular resolutions down to a few hundreds of au. The Galactic center is a unique region of our Milky Way in terms of star formation, stellar density, turbulence and temperature of the ISM, the strong magnetic fields, and large number of stellar remnants. It is an ideal region to probe the physics of galactic nuclei in much detailed fashion than can be achieved in extragalactic galactic nuclei.

\section{Magnetic fields in star formation}

\cite{Bourke01.2026.SKA} looks into the SKAO capability to determine the magnetic field strengths (B) in molecular clouds and star forming regions, down to protoplanetary disks using the Zeeman effect (splitting of energy levels, directly proportional to B) in spectral lines (atomic H~\textsc{I}, molecules such as OH, CH, CCS, but also in RRLs and masers). Until now, detections are rare in the diffuse clouds in view of the low strength, from $\mu$G to few tens of mG. However, such measurements are crucial to assess the importance of magnetic fields in the star formation process, as the few current observations suggest marginally supercritical conditions where the magnetic fields cannot prevent collapse but inhibit it, but it remains unclear if this is a biased result. A large-scale, statistical investigation with SKAO can, thus, lead to a significant advancement in our understanding of the star formation process. 

\citet{Bracco01.2026.SKA} also investigates the importance of magnetic fields in the star formation process, but focuses on synchrotron emission in dense starless cores. This early phase of star formation is critical for understanding how mass is accreted during star formation. Complementary to the Zeeman splitting method (sensitive to the the line-of-sight B) and the dust polarization (sensitive to the plane-of-the-sky B, i.e., perpendicular to the line of sight), synchrotron emission provides an indirect probe of B. While electrons can be thermally distributed, the presence of non-thermal electrons is of great interest as it depends on the the penetration of Galactic cosmic rays into molecular clouds. The radio regime is ideal to probe the optically thin regime of synchrotron emission ($\nu^{-0.7}$), but only few previous measurements have been obtained so far in dense clouds. Both SKA-Low and SKA-Mid can, thus provide information on the surface brightness and spectral index in magnetised dense starless cores.

\section{Resolving the origin of the Anomalous Microwave Emission}

\citet{Vidal01.2026.SKA} address a peculiar excess of radiation in the $10-60$~GHz range, distinct from synchrotron, free-free, or thermal dust emission, called the Anomalous Microwave Emission (AME). It has been detected in diverse conditions, from diffuse ISM clouds to protoplanetary disks and even in galaxies, and it is, therefore, a significant additional foreground component to cosmic microwave background studies. The common explanation for the AME is electric dipole radiation from rapidly rotating (spinning) small dust grains, but other mechanisms were propsed such as  magnetic dipole emission (MDE). SKAO can help map the AME morphology and spectral energy distribution throughout our Galaxy, and disentangle the different models, thus probing  interstellar grain physics and the small-scale structure of the ISM.

\section{Magnetic activity, magnetospheres, and stellar endpoints}

SKAO will probe the magnetic fields in stars of different masses, ages, and stellar evolution stages:

\citet{Cavallaro01.2026.SKA} provide an overview of SKAO's capabilities to investigate the coherent and incoherent radio emission from magnetospheres in stars, from low-mass stars, ultracool dwarfs to massive A/B-type stars. Incoherent radio emission is generally explained by the centrifugal breakout (CBO) mechanism in massive stars, where non-thermal electrons produce incoherent gyro-synchrotron emission. The radio luminosity is found to follow the power of the CBO events. Interestingly, a similar trend in radio luminosity exists in ultracool dwarfs, despite the lack of strong winds, suggesting nevertheless a similar process. In addition, both types of stars exhibit coherent electron cyclotron maser emission that is polarised and rotationally modulated, in a similar fashion as observed in aurorae in Solar system planets. The wide area covered by SKAO will, thus, offer the opportunity to detect many stellar magnetospheres, and thus study the CBO mechanism throughout our Galaxy.

\citet{LorenAnderson01.2026.SKA} investigate the advancements that SKAO will allow to achieve in the field of massive stars and stellar clusters, which provide strong feedback mechanisms into the ISM, such as radation pressure, photoionisation, winds, and cosmic ray acceleration. Radio observations can trace the free-free emission from ionised gas and non-thermal synchrotron
emission in shocks. Physical conditions can be probed via RRLs (hydrogen, helium, carbon), including their Zeeman splitting, and synchrotron emission. SKAO will provide strong constraints on the stellar wind mass-loss
rates, on the photoionized gas kinematics and dynamics, and will explore photodissociation regions around ultracompact H~\textsc{ii} regions. This will provide crucial insights into feedback mechanisms in our Galaxy and beyond.

\citet{Buemi01.2026.SKA} focus on the study of the  stellar winds and circumstellar environment of evolved massive stars. In the transitional phases (e.g., Luminous Blue Variable, Wolf Rayet, Red Supergiants) toward supernova explosion, mass loss (steady or eruptive) significantly disturbs the surrounding environment, which will significantly affect the future signatures of supernovae. SKAO will provide a detailed characterisation of the circumstellar environment, which can then help us quantify the mass-loss phenomena and will, thus, lead us to a better understanding of the connections of core-collapse supernovae   and remnants, thanks in part to the combination of radio observations with magnetohydrodynamical simulations and complementary multi-wavelength data. 

\citet{Ingallinera01.2026.SKA} investigated the last stages of massive stellar evolution, namely supernova remnants (SNR). Despite the long history of radio observations of SNR, questions still remain, in particular their contributions to Galactic cosmic rays. SKAO will make a significant step in SNR studies, beyond SKA precursors, by providing deep sensitivities to faint SNR in polarisation, revealing their detailed structures, and their magnetic fields. The exquisite angular resolution will further help in tracing fine structures such as filaments and shock fronts, while the wide spectral coverage will provide access to spectral turnovers and breaks or cut-offs, making the link between radio and X/$\gamma$-ray emission and, thus, modeling the non-thermal emission.

\section{Masers in our Galaxy}

\citet{Rygl01.2026.SKA} provided an interesting overview of the maser emission in different sources in our Galaxy, and its capability to probe physical processes, e.g., revealing the dynamics in protostars and their disks, tracing winds and envelopes, mapping the Galactic structure, etc. Masers are pumped via collisions or radiation, and are good tracers of the gas and dust density and temperature. They are cosmic rulers since they are intrinsically compact and bright, providing access to dense regions otherwise generally inaccessible, from a few au to kpc, thanks to their astrometric capabilities. Masers in different molecules (OH, CH$_3$OH, C$_2$H$_2$) are available with SKA-Mid, while discoveries of new types of masers are expected. The chapter provides an overview of maser capability in star formation and evolution, demonstrating its potential for the science verification and calibration phases of SKAO.

\section{Conclusions}

The chapters linked to the OG Science Working Group show the breadth of science that SKAO will address in the diffuse and stellar components of our Galaxy. 
The 13 chapters represent current interests of the OG community in the context of the 14 SKA science working groups. Clearly, chapters covered by other SWG pertain to OG. Furthermore, synergies between the OG chapters and other SWG are highlighted as follows:
\begin{itemize}
    \item The Magnetism SWG has important synergy with chapters focused on magnetic fields \citep[e.g.][]{Bourke01.2026.SKA,Bracco01.2026.SKA,Yamamoto01.2026.SKA}. We also point out the chapter of \citet{Robishaw01.2026.SKA}, focusing on measuring magnetic Fields via the Zeeman effect, in both our Galaxy and nearby galaxies, providing further synergies between the OG, Magnetism, Cradle of Life, and Extragalactic SWGs.
    \item The Cradle of Life SWG has potential synergy with the Galactic center chapter \citep{Schoedel01.2026.SKA} and possibly the spectroscopic surveys chapter \citep{Karska01.2026.SKA}, due to the possible detection of new and prebiotic molecules.
    \item As the nearest and best studied galactic nucleus, the Galactic center \citep{Schoedel01.2026.SKA} is of fundamental interest to the extragalactic community.
    \item Monitoring of maser variability \citep{Rygl01.2026.SKA} provides commensal data that is potentially useful for the Transient and Extragalactic working groups, e.g., by comparing the Galactic structure with that of nearby resolved spiral galaxies. 
    Commensality can also be provided by the Galactic plane survey \citep{Traficante01.2026.SKA} and observations of supernovae remnants \citep{Ingallinera01.2026.SKA}.
    \item Opportunities for science verification (SV) is discussed in the maser chapter \citep{Rygl01.2026.SKA}. SV observations can also be a useful pilot part of some chapters including the Galactic plane survey \citep{Traficante01.2026.SKA}, spectroscopic surveys \citep{Karska01.2026.SKA}, and the Galactic center \citep{Schoedel01.2026.SKA}, owning to the nature of the proposed observations.
\end{itemize}

Table 1 summarizes the SKA capabilities requested by the OG chapters. The OG community has an extremely high demand for SKA-Mid in AA4 configuration: all chapters requested SKA-Mid, and all but one chapter requested AA4. About half of the chapters (7/13) discussed reduced yet useful capability of AA* for their proposed scientific goals. SKA-Low is only requested in 3 chapters. Few chapters explored capabilities beyond AA4 \citep[e.g.][]{Salas01.2026.SKA,Bourke01.2026.SKA,Ingallinera01.2026.SKA}.

In summary, SKAO will provide crucial, transformative new results that will help us understand how matter cycles in the ISM into molecular clouds, and how stars are formed in magnetised molecular clouds and cores, and how they shape their circumstellar environment in the later phases of stellar evolution toward supernova explosions. The OG community is excited and eager to observe our Milky Way with SKAO in the coming years.

\begin{table}
\begin{center}
\caption[]{Summary of required SKA capabilities by chapter}
\setlength\tabcolsep{2pt} 
\begin{tabular}{lcccccc}
\hline \noalign {\smallskip}
Chapter/topic & SKA-Low & SKA-Mid & AA* & AA4 & Beyond AA4 & Section\\
\hline
  Yamamoto et al. (2026), HI-H$_2$ &  & Y &  & Y &  & 2\\
  Salas et al. (2026), RRLs & Y & Y & Y & Y &  Y& 2\\
  Traficante et al. (2026), GP survey &  & Y & Y & Y &  & 2\\
  Karska et al. (2026), Spec. survey &  & Y &  & Y &  & 2\\
  Schoedel et al. (2026), Galactic Center &  & Y &  & Y &  & 2\\
  Bourke et al. (2026), Zeeman &  & Y & Y & Y & Y & 3\\
  Bracco et al. (2026), Starless cores & Y & Y & Y & Y &  & 3\\
  Vidal et al. (2026), AME &  & Y & Y &  &  & 4\\
  Cavallaro et al. (2026), Magnetospheres &  & Y &  & Y &  & 5\\
  Andersson et al. (2026), Stars \& clusters &  & Y &  & Y &  & 5\\
  Buemi et al. (2026), Evolved stars &  & Y & Y & Y &  & 5\\
  Ingallinera et al. (2026), SNRs & Y & Y & Y & Y & Y & 5\\
  Rygl et al. (2026), Masers &  & Y &  & Y &  & 6\\
\hline \noalign {\smallskip}
  Count & 3 & 13 & 7 & 12 & 3 & \\
  Percentage & 23\% & 100\% & 54\% & 92\% & 23\% & \\
\hline\end{tabular}
\end{center}
\end{table}

\section*{Acknowledgments}
This work is partly supported by the National SKA Program of China (2025SKA0140100).

\bibliographystyle{abbrvnat-maxbibnames4}
\bibliography{chapter} 

\end{document}